\documentclass[aps,prd,showpacs,showkeys,superscriptaddress,twocolumn]{revtex4-1}
\usepackage{amsmath}
\usepackage{amssymb}
\usepackage{graphicx}
\usepackage{dcolumn}
\usepackage{bm}
\usepackage{color}
\usepackage{slashed}
\usepackage{lipsum} 
\usepackage{amsmath} 
\allowdisplaybreaks[4] 

\begin{document}

\preprint{APS/123-QED}

\title{Transverse Ward-Takahashi Identities and Full Vertex Functions in Different Representations of QED$_3$}
\thanks{Manuscript submitted to Chin. Phys. C}

\author{Cui-Bai Luo}
 \email{cuibailuo@ahnu.edu.cn}
  \affiliation{Department of Physics, Anhui Normal University, Anhui 241002, China} 
  \affiliation{Department of Physics, Nanjing University, Nanjing 210093, China}
 
\author{Hong-Shi Zong} 
 \email{zonghs@nju.edu.cn}
   \affiliation{Department of Physics, Nanjing University, Nanjing 210093, China}
   \affiliation{Department of Physics, Anhui Normal University, Anhui 241002, China} 
   \affiliation{Nanjing Proton Source Research and Design Center, Nanjing 210093, China}



\date{\today}

\begin{abstract}
We firstly derive the transverse Ward-Takahashi identities (WTI) of $N-$dimensional quantum electrodynamics by means of the canonical quantization method and the path integration method, and then try to prove that QED$_3$ is solvable based on the transverse WTI and the longitudinal WTI, that is, the full vector and tensor vertices functions can be expressed in term of the fermion propagators in QED$_3$. Further, we discuss the effect of different $\gamma$ matrix representations on the full vertex function.
\begin{description}

\item[Usage]
 
\end{description}
\end{abstract}

\maketitle



\section{Introduction }
The normal (longitudinal) Ward-Takahashi identities (WTI) \cite{ward} play an important role in various problems in quantum field theory, for example, it provides a consistency condition in the perturbative and non-perturbative approach of any quantum field theory and the proof of renormalizability of gauge theories \cite{proof}. In the Dyson-Schwinger equations (DSEs) approach, the fermion-boson vertex function is an important quantity to be specified.  In order to use DSEs to do some actual calculations, we must artificially cut off the coupling of the \textit{n}-point Green's and the higher-order Green's function to properly close the DSEs. If one can express the three-point vertices in terms of the two-point functions, the DSEs which consists of an infinite set of coupled integral equations will form a closed system for the two-point functions. Among the many vertex approximations, the most famous one is the  Ball-Chiu Ansatz \cite{ballchiu} except for the bare vertex approximation. How to properly break through the bare vertex approximation and the  Ball-Chiu Ansatz is a very challenging subject. 

In what follows, we try to do some work in this area to see what kind of conditions can make the DSEs closed. One possible approach to this problem is to use the WTI to constrain the form of the vertex function. However, the normal WTI only contains the longitudinal part of the vertex functions, leaving its transverse part undetermined. In order to find further constraints on the vertex function, Takahashi derived so called transverse relations \cite{twardys} relating Green’s functions of different orders to complement the normal WTI, which have the potential to determine the full fermion-boson vertex in terms of the renormalization functions of the fermion propagator \cite{tk}. Subsequently,  He, Takahashi \cite{tward3, tward2} and Kondo \cite{tward}, \textit{etc}.,  find that the complete set of transverse WTI and longitudinal WTI for the vector, axial-vector and tensor vertex functions can form the complete solutions for these vertex functions in four-dimensions gauge theories. When it ignore the contribution of the three integral-term involving the Wilson line and chiral limit $m \to 0$, the full vector vertex functions are expressed in terms of the two-point functions. Subsequently,  Pennington and R. Williams \cite{penwill, twardzx, tward8} checked the transverse WT identity for the fermion-boson vertex to one-loop order. 

Some authors also attempt to study this problem in other various ways such as via constraints \cite{wt1} and direct numerical solution \cite{wt2}. For instance, Qin \textit{etc.,} \cite{qin} consider the coupling of a dressed-fermion to an Abelian gauge boson, and describe a unified treatment and solution of the longitudinal and transverse WTI. The vector vertex is discussed by using twelve independent tensor structures. What we need to emphasize here is that although people have made important progress in constructing the fermion-boson vertex functions, it is still not possible to represent the full vector, axial-vector and tensor vertex functions in the four-dimensional gauge theory by two-point functions. That is to say, in four-dimensional quantum electrodynamics (QED), one cannot construct a completely closed DSEs by three-point Green`s functions and two-point Green`s functions.

However, when the dimension of the gauge theory is reduced, it will change a lot. For example, in the case of two-dimensional gauge theory, Kondo first pointed out in Ref. \cite{tward} that ``the transverse together with the usual (longitudinal) Ward-Takahashi identity are applied to specify the fermion-boson vertex function. ...... It is especially shown that in two dimensions, it becomes the exact and closed Schwinger-Dyson equation which can be exactly solved''. Since QED$_3$ can be regarded as an effective theory of high temperature superconductivity or as a toy model of quantum chromodynamics (QCD), this makes the research on QED$_3$ especially interesting (related can be found in the relevant literature in Ref. \cite{xia}). So a very natural question arises, can people get a completely Closed DSEs in three-dimensional QED (QED$_3$)?  

The main purpose of this work is to try to construct a closed DSEs by three-point and two-point Green's functions in the case of QED$_3$, based on normal and transverse WTI. In addition, given that there are two different expressions of the $\gamma$ matrix in QED$_3$, we will also discuss the effect of different $\gamma$ matrix representations on the full vertex functions.

\section{Full Vertices Functions}\label{sec2}
\subsection{Vertex Functions in \textit{N}-dimensional Gauge Theory}
The longitudinal (normal) WT identity determines its divergence, \textit{i.e.}, $\partial_\mu \Gamma^\mu (x; y,z) $. The transverse WT identity \cite{twardys} specifies the curl of the vertex function $\partial^\mu \Gamma^\nu (x; y,z) - \partial^\nu \Gamma^\mu(x;y,z)$, where $\Gamma^\mu (x;y,z)$ is the fermion-boson (photon) vertex function. It was derived by Takahashi in 1986. The transverse Ward-Takahashi identity can be converted to
\begin{flalign}
&\partial_x^\mu \langle 0|Tj (x) \psi(y) \bar{\psi}(z)|0\rangle - \partial_x^\nu \langle 0|Tj (x) \psi(y) \bar{\psi}(z)|0\rangle, & 
\end{flalign}
where $j(x)$ is the current operators. The above relation is valid for both QED and QCD.

Firstly one introduces two bilinear covariant current operators, 
\begin{flalign} \label{bilinear}
&V^{\rho \mu \nu \lambda} (x)=\frac{1}{4} \bar{\psi}(x) \bigg[[\gamma^\rho, \sigma^{\mu \nu}], \gamma^\lambda  \bigg] \psi(x) =g^{\rho \mu} j^{\nu \lambda}(x) -g^{\rho \nu} j^{\mu \lambda}(x), & \nonumber  \\
&V^{\rho \mu \nu} (x) =\frac{-i}{2} \bar{\psi}  [\gamma^\rho, \sigma^{\mu \nu}] \psi =g^{\rho \mu} j^\nu(x) -g^{\rho \nu} j^\mu(x) .& 
\end{flalign}
One needs to calculate the curl of the time-ordered products of the fermion's three point functions involving the vector, axial-vector and tensor current operators, namely $j^\mu(x) =\bar{\psi}(x)\gamma^\mu \psi(x), j^{\mu \nu}(x) =\bar{\psi}(x)\sigma^{\mu\nu} \psi(x)$ and $ j_5^\mu(x) =\bar{\psi}(x)\gamma^\mu \gamma_5 \psi(x)$, respectively. Then the transverse WTI for fermion's vertex functions can be obtained by the curl of the $T$ products of the corresponding fermion's three-point function
\begin{flalign}
&\partial_\rho^x \langle 0| TV^{\rho \mu \nu (\lambda)} (x)\psi(y) \bar{\psi}(z) |0\rangle &\nonumber \\
&=\partial_x^\mu \langle 0|Tj^{\nu (\lambda)} (x)\psi(y) \bar{\psi}(z)|0\rangle  - \partial_x^\nu \langle 0|Tj^{\mu (\lambda)}  (x) \psi(y) \bar{\psi}(z)|0\rangle.&
\end{flalign}

For the convenience of the discussion in \textit{N}-dimensional gauge theory, we only use the relations of gamma matrices that does not depend  on the space–time dimensions and does not introduce $\gamma_5$ matrix, which can be expressed as follows
\begin{flalign} \label{ngamma}
&\{\gamma^\mu, \gamma^\nu\}_N =2g^{\mu \nu}, \quad \frac{i}{2}[\gamma^\mu, \gamma^\nu] _N= \sigma^{\mu\nu}, &\nonumber \\
&\frac{1}{2}[\gamma^\rho, \sigma^{\mu \nu}]_N = i[g^{\rho \mu}\gamma^\nu - g^{\rho \nu}\gamma^\mu]. &
\end{flalign}

There are two ways to compute the curl of the time-ordered products of the above three-point functions, one is the canonical quantization method, and the other is the path integration method. Please refer to the Appendix for the derivation. Through the canonical quantization and path integration method, we arrive at the transverse WT relations for the fermion’s vertex functions in \textit{N}-dimensional gauge theory in configuration space, 
\begin{flalign} \label{vector current 1}
& \partial^\mu   \langle 0| T j^\nu (x)  \psi(y) \bar{\psi}(z)|0\rangle  - \partial^\nu   \langle 0| T j^\mu (x) \psi(y) \bar{\psi}(z)|0\rangle &\nonumber \\
&= \lim_{x'\to x}( \partial^{x'}_\rho - \partial^{x}_\rho)\langle 0|  T \bar{\psi}(x')\frac{i}{2}\{\gamma^\rho, \sigma^{\mu \nu}\}  U(x',x)\psi(x) \psi(y)\bar{\psi}(z) |0\rangle &\nonumber \\ 
&+i \sigma^{\mu \nu}  \delta^4(x-y) \langle 0|   T \psi(x)  \bar{\psi}(z)  |0\rangle  &\nonumber \\
&+ i\langle 0|  T\psi(y)  \bar{\psi}(x)   0\rangle  \sigma^{\mu \nu }  \delta^4(x-z) & \nonumber \\
&+2m   \langle 0| T \bar{\psi}(x)   \sigma^{\mu \nu} \psi(x)  \psi(y)\bar{\psi}(z) |0\rangle &  
\end{flalign}
and
\begin{flalign} \label{tensor current 1}
& \partial^\mu   \langle 0| T j^{\nu \lambda}(x)  \psi(y) \bar{\psi}(z)|0\rangle  - \partial^\nu   \langle 0| T j^{\mu \lambda} (x) \psi(y) \bar{\psi}(z)|0\rangle& \nonumber \\
&= -\frac{1}{2}\{\sigma^{\mu \nu} , \gamma^\lambda\} \delta^4(x-y)   \langle 0| T\psi(x) \bar{\psi}(z) |0\rangle & \nonumber \\
&+\langle 0|T  \psi(y) \bar{\psi}(x) | 0\rangle \frac{1}{2}\{\sigma^{\mu \nu} , \gamma^\lambda\} \delta^4(x-z) &\nonumber \\
&-  (\partial_\rho^{x'}-\partial_\rho^{x})  \langle 0|  T\bar{\psi}(x') \frac{1}{4}\bigg[  \gamma^\rho, \bigg\{ \sigma^{\mu \nu}, \gamma^\lambda  \bigg\}\bigg] U(x',x)\psi(x)  \psi(y) \bar{\psi}(z) | 0\rangle & \nonumber \\
 &-(\partial^{\lambda(x')}+ \partial^{\lambda(x)})\langle 0|  T\bar{\psi}(x')\sigma^{\mu \nu  } U(x',x)\psi(x)  \psi(y) \bar{\psi}(z) | 0\rangle. &
\end{flalign}

\subsection{The Anomaly}
The symmetry of the classical theory may be destroyed by quantum anomaly and there is the corresponding anomalous WT identity \cite{abj1}, this must be considered in advance when studying the full vertex functions. In four-dimensional gauge theories, by using perturbative method and Pauli-Villars regularization and dimensional regularization, Sun, \textit{et al.}, \cite{tward4} find that there is no transverse anomaly term for both the axial-vector and vector current. The absence of transverse anomalies for both axial-vector current in QED$_2$ theory and vector (tensor) current in QED$_3$ theory are also verified respectively \cite{cbl}. So in the case of transverse WT identity, one don't need to discuss the problem of transverse quantum anomalies. However, the quantum anomaly of longitudinal WTI  needs to draw our attention. 

\subsection{Representation and Full Vertices Functions}
In the above, we established the relationships of transverse WTI (\ref{vector current 1}, \ref{tensor current 1}) using only matrix relations (\ref{ngamma}), which is suitable for $N-$dimensional time-space.  As we will see shortly, the representation of symmetrized part $\{\gamma^\rho,\sigma^{\mu \nu} \}$ depends on the space–time dimensions. In $3+1$ dimensions time-space, substituting $\{\gamma^\rho,\sigma^{\mu \nu} \}=-2\epsilon^{\rho \mu \nu \lambda}\gamma_\lambda \gamma_5$ into transverse WTI (\ref{vector current 1}, \ref{tensor current 1}), it is easy to find that our results are exactly the same as those given in Ref. \cite{tward3}. This can be seen as a self examination of the transverse WTI (\ref{vector current 1}, \ref{tensor current 1}).

Now we turn to consider the $2+1$ dimensional case, we choose the following gamma matrices
\begin{flalign}\label{3gamma}
&\gamma^0 = \sigma^3, \quad  \gamma^1 = i\sigma^1, \quad \gamma^2 = i\sigma^2,\quad \frac{1}{2}\{\gamma^\rho,\sigma^{\mu \nu} \}= \epsilon^{\rho \mu \nu },&
\end{flalign}
where $\sigma^i$ denotes Pauli matrix.  In this case, we don't have the freedom to construct additional gamma matrices that anti-commute with all $\gamma^\mu$ in the  $2 \times 2$ representation. This means that the flavor symmetry of fermions is the same whether they have mass or not.

Substituting relations (\ref{3gamma}) into Eqs. (\ref{vector current 1}, \ref{tensor current 1}), the transverse Ward-Takahashi identity for the vector and the tensor   vertex can be written in momentum space by introducing the standard definition for the three-point function,
 \begin{flalign}  \label{vector} 
&q^{\mu } \Gamma^{\nu }_V  (p_1,p_2) -  q^{\nu } \Gamma^{\mu }_V  (p_1,p_2)& \nonumber \\
=& -i S^{-1}_F(p_1) \sigma^{ \mu \nu } -  i\sigma^{\mu \nu }  S^{-1}_F(p_2)+i \epsilon^{ \rho \mu \nu} ( p_{1\rho}+p_{2\rho })   \Gamma_S(p_1,p_2) & \nonumber \\
&  - 2im \Gamma_{T}^{\mu \nu}(p_1,p_2)-i \int \frac{d^3k}{(2\pi)^3}2k_\rho \epsilon^{ \rho \mu \nu} \Gamma_S(p_1,p_2,k) &
 \end{flalign}
and
\begin{flalign} \label{tensor}
& q^{\mu}  \Gamma^{\nu \lambda}_T(p_1,p_2) - q^\nu   \Gamma^{\mu \lambda}_T(p_1,p_2) +q^\lambda     \Gamma^{\mu \nu}_T(p_1,p_2) &\nonumber \\  
&= \epsilon^{\mu \nu \lambda}S^{-1}_F(p_1) -  \epsilon^{\mu \nu \lambda} S^{-1}_F(p_2),&
\end{flalign}
where $\Gamma_S, \Gamma^\mu_V, \Gamma^{\mu \nu}_T$ are the scalar, vector and tensor vertex functions, respectively,  and $q=(p_1-k)-(p_2-k)$.  The last term in Eq. (\ref{vector}) is called the integral-term involving the vertex function $\Gamma_{S}(p_1,p_2; k)$ with the internal momentum $k$ of the gauge boson appearing in the Wilson line, which is defined by the Fourier transformation
\begin{flalign}  
&\int d^3x d^3 x' d^3x_1d^3x_2 \langle 0| T\bar{\psi}(x)\psi(x)\bar{\psi}(x_1)\psi(x_2) U(x',x)|0\rangle \cdot &\nonumber \\ 
& e^{i(p_1 x-p_2 x_2 -(p_1-k)x' +(p_2-k)x )}& \nonumber \\
=&(2\pi)^3\delta^3(p_1-p_2-q)iS_F(p_1)\Gamma_S(p_1,p_2,k)iS_F(p_2).& \nonumber 
\end{flalign}
The integral-term to one-loop order in four dimensional gauge theory have been calculated in Ref. \cite{twardzx}.  

Noting that if one chooses the basic fermion field to be a four component spinor, the three $4 \times 4$ $\gamma$ matrices can be taken to be
\begin{flalign}
& \gamma^0 =\left[
 \begin{matrix}
\sigma_3 &0  \\
 0&- \sigma_3
  \end{matrix}
  \right], 
\gamma^1 =i\left[
 \begin{matrix}
\sigma_1 &0  \\
 0& - \sigma_1
  \end{matrix}
  \right], 
 \gamma^2 =i\left[
 \begin{matrix}
\sigma_2 &0  \\
 0&- \sigma_2
  \end{matrix}
  \right],& 
\end{flalign}
where we can define a $4\times4$ matrices $\gamma^5$ that anti-commute with all $\gamma^\mu$
\begin{flalign}
& \gamma^3 =i\left[
 \begin{matrix}
0 &\mathnormal{I} \\
 \mathnormal{I}& 0 
  \end{matrix}
  \right], 
 \gamma^5=i\left[
 \begin{matrix}
0&\mathnormal{I} \\
-\mathnormal{I}&0
  \end{matrix}
  \right].&
\end{flalign}
This is different to the $2\times 2$ representation where there is no $\gamma_5$ matrix and dynamical chiral symmetry breaking. Since such differences in symmetry are expected to equally manifest in the vertices as well as the propagators. It can be expected that the vertices cannot be equal in these different representations.

At this point we have
\begin{flalign}
&\frac{1}{2}\{\gamma^\rho,\sigma^{\mu \nu} \}= \epsilon^{\rho \mu \nu }\gamma_M, \quad \gamma_M = \left[
 \begin{matrix}
\mathnormal{I}&0 \\
0&-\mathnormal{I}
  \end{matrix}
  \right], \quad [\gamma_M, \gamma^\rho]=0.&
\end{flalign}
So in this case, through similar derivation steps, the above relation Eqs. (\ref{vector}, \ref{tensor}) will be modified as follows:
\begin{flalign}  \label{vector2} 
&q^{\mu } \Gamma^{\nu }_V  (p_1,p_2) -  q^{\nu } \Gamma^{\mu }_V  (p_1,p_2)& \nonumber \\
&= i \epsilon^{ \rho \mu \nu} ( p_{1\rho}+p_{2\rho })   \Gamma_M(p_1,p_2)-i \int \frac{d^3k}{(2\pi)^3}2k_\rho \epsilon^{ \rho \mu \nu} \Gamma_M(p_1,p_2, k) & \nonumber \\
&-i S^{-1}_F(p_1) \sigma^{ \mu \nu } -  i\sigma^{\mu \nu }  S^{-1}_F(p_2)  - 2im \Gamma_{T}^{\mu \nu}(p_1,p_2)& 
 \end{flalign} 
and
\begin{flalign} \label{tensor2}
& q^{\mu}  \Gamma^{\nu \lambda}_T(p_1,p_2) - q^\nu   \Gamma^{\mu \lambda}_T(p_1,p_2) +q^\lambda     \Gamma^{\mu \nu}_T(p_1,p_2) &\nonumber \\  
&= \epsilon^{\mu \nu \lambda}S^{-1}_F(p_1) \gamma_M -  \epsilon^{\mu \nu \lambda} \gamma_M S^{-1}_F(p_2),&
\end{flalign}
where $\Gamma_M$ denotes vertex function $ \langle 0|  T \bar{\psi}(x)\gamma_M\psi(x) \psi(y)\bar{\psi}(z) |0\rangle $ in momentum space. Compared Eqs.(\ref{vector2}, \ref{tensor2}) with the Eqs.~(\ref{vector}, \ref{tensor}), it is found that the full vertex function does depend on the different $\gamma$ matrix representation we use.

The above Eqs. (\ref{tensor}, \ref{vector}) show that the transverse part of the vertex function is related to the inverse of the fermion propagator and other vertex functions, namely the full vertex functions are coupled to each other and form a set of coupled equations. For instance, Eq. (\ref{vector}) shows that the transverse part of the vector vertex function is related to the fermion propagator, the tensor and scalar vertex functions. Noting that the transverse WT relation for the tensor vertex functions in four-dimensions have a psudoscalar vertex functions term \cite{tward3}, which is different from the case of three-dimensions (see Eqs. (\ref{tensor}) and (\ref{tensor2})). The reason why the result in three-dimensions is quite different from that in four-dimensions is due to following facts: 
\paragraph{$\gamma$ matrices} The $\gamma$ matrices representation in QED$_3$ is different from that in QED$_4$ and the commutative relations of $\gamma$ matrices are also different (see the Eq. (\ref{tensor current}), where $\bigg[  \gamma^\rho, \bigg\{ \sigma^{\mu \nu}, \gamma^\lambda  \bigg\}\bigg] =0 $ in QED$_3$, but not in QED$_4$ ), which leads to the transverse Ward-Takahashi identities in QED$_3$ for vertex functions to be simpler; 

\paragraph{Integral-term involving the vertex function } In QED$_4$, the transverse  Ward-Takahashi identities for the vector vertex function contains the integral-term involving the axial-vector vertex function, but the axial-vector vertex function cannot be expressed by the two-point Green's function. However, in QED$_3$, the vector vertex function contains the integral-term involving the scalar vertex function $q_\nu \int \frac{d^3k}{(2\pi)^3}2k_\rho \epsilon^{ \rho \mu \nu} \Gamma_S(p_1,p_2, k)$, while the scalar vertex function $\Gamma_S$ can be expressed by two-point Green's function (please refer to the Eq. (\ref{b2}) ), due to the antisymmetry of $\epsilon^{ \rho \mu \nu}$ and $\sigma^{ \mu \nu }$.

For the above reasons, we don't need to make any approximation in the current work to get a completely closed DSEs in QED$_3$, which is the biggest difference between our present work and the past works. Now we begin to derive the complete expression of the vertex function. 

The well-known normal Ward-Takahashi identities 
\begin{flalign}\label{vector normal}
&q_\mu \Gamma^{\mu }_{V  } (p_1,p_2) =  S^{-1}_F(p_1) -S^{-1}_F(p_2) &\nonumber \\
&iq_\mu \Gamma^{\nu \mu }_{T  } (p_1,p_2) = S^{-1}_F (p_1) \gamma^\nu +\gamma^\nu S^{-1}_F(p_2) + 2m\Gamma_V^\nu(p_1,p_2)  & \nonumber \\
&+ (p_1^\nu + p_2^\nu)\Gamma_S(p_1,p_2) &
\end{flalign}
denote the longitudinal part of the three-point vertex function, which together with the transverse WT relation form a complete set of WT-type constraint relations for the fermion's three-point vertex functions in QED$_3$ theories. Then by this complete set of constraint relations one can obtain the complete solutions for these vector and tensor vertex functions. 

Obviously, in four-dimensions space-time, it is extremely difficult to consider the full contributions of the above three Wilson integral-terms in Eqs. (\ref{vector current 1}, \ref{tensor current 1}). In order to get a set of closed DSEs,  He \cite{tward3, tward2} firstly ignore the contribution of integral-term involving the vertex functions and see what will happen. However,  in three-dimensions space-time, we find that the full vertex function has a very simple expression (no need to ignore the integral term), which can be expressed in terms of the fermion propagators. By using Eqs. (\ref{tensor}, \ref{vector}) and normal Ward-Takahashi identities Eq. (\ref{vector normal}), the complete expression in $2\times 2$  representation for the vector vertex can be obtained as following
\begin{flalign}\label{vector later}
&\Gamma_V^\mu (p_1,p_2) = \frac{1}{ ( q^{2}-4m^2) }\bigg\{ q^{\mu }    \bigg[S^{-1}_F(p_1)  -    S^{-1}_F(p_2)\bigg]  & \nonumber \\
  &+ i   q_{\nu } \bigg[S^{-1}_F(p_1) \sigma^{ \mu \nu } +   \sigma^{\mu \nu }  S^{-1}_F(p_2) \bigg] &\nonumber \\
  &+2 m   \bigg[ S^{-1}_F(p_1) \gamma^\mu + \gamma^\mu S^{-1}_F(p_2) \bigg] &\nonumber \\
& + \bigg[2 m (p_1^\mu+ p_2^\mu)    -i \epsilon^{ \rho \mu \nu}  q_\nu ( p_{1\rho}+p_{2\rho }) \bigg]  \Gamma_S(p_1,p_2)& \nonumber \\
&+i \int \frac{d^3k}{(2\pi)^3}2k_\rho q_\nu \epsilon^{ \rho \mu \nu} \Gamma_S(p_1,p_2,k) \bigg\}.&
  \end{flalign} 
And the tensor vertex function is
\begin{flalign}\label{tensor later}
&q^2\Gamma^{\mu \nu}_{T}(p_1,p_2) =i\bigg\{S_F^{-1}(p_1)(q^{\mu}\gamma^\nu - q^\nu \gamma^\mu-iq_\lambda \epsilon^{\mu \nu \lambda}) &\nonumber \\
&+ (q^{\mu}\gamma^\nu -q^\nu \gamma^\mu + iq_\lambda \epsilon^{\mu \nu \lambda}   ) S_F^{-1}(p_2)  +2m[q^\mu\Gamma^\nu_V(p_1,p_2) &\nonumber \\
&-q^\nu\Gamma^\mu_V(p_1,p_2) ] +[q^\mu (p_1^\nu +p_2^\nu) -  q^\nu (p_1^\mu +p_2^\mu) ]\Gamma_S(p_1,p_2)\bigg\},&
\end{flalign}
where
\begin{flalign}\label{b2}
q_\mu (p_1^\mu+ p_2^\mu)\Gamma_S(p_1,p_2) &=-2m\bigg[S^{-1}_F(p_1) -S^{-1}_F(p_2)\bigg] &\nonumber \\
&-\bigg[ S^{-1}_F(p_1) \gamma^\mu q_\mu  + \gamma^\mu q_\mu S^{-1}_F(p_2) \bigg]. &
 \end{flalign} 
 
It is very interesting here to examine the possible kinematic singularities that the dressed vertex function Eqs. (\ref{vector later}-\ref{bl}) may have. In the case of the chiral limit, we have
\begin{flalign}\label{bl}
\Gamma_S(p_1,p_2)  &=  \frac{-1}{(p_1^\mu+ p_2^\mu) } \times  \bigg[ S^{-1}_F(p_1) \gamma^\mu  + \gamma^\mu S^{-1}_F(p_2) \bigg],&
 \end{flalign} 
there is no singularity for $\Gamma_S(p_1,p_2) $ [in the limit $q_\mu \to 0$ (which requires both $q^2 \to 0$ and $p^2_1 \to p^2_2$ in Minkowski metric. It is also a limit in Euclidean space) and the limit $q^2 \to 0$ for $p_1^2 \neq p_2^2$ in Minkowski metric]. In the case of non-Chiral limit, there is also no singularity for $\Gamma_S(p_1,p_2)$ in the limit $q^2 \to 0$ for $p_1^2 \neq p_2^2$ in Minkowski metric. So the vector vertex function $\Gamma^{\mu}_V$  does not suffer from singularities in the limit $q^2 \to 0$ for $p_1^2 \neq p_2^2$, due to $q_\mu \neq 0$. However, the tensor vertex function $\Gamma^{\mu \nu}_{T}(p_1, p_2)$ always has singularity. This singularity is worthy of careful consideration.

The full vertex functions depend on the different $\gamma$ matrix representation we use. If we replace $\Gamma_S$ with $\Gamma_M$, in the above equation (\ref{vector later}) we get the vector vertex function in $4\times 4$ representation. Similarly, by replacing $\epsilon^{ \mu \nu \lambda}$ with $\epsilon^{ \mu \nu \lambda}\gamma_M$ in the equation (\ref{tensor later}) subsequently we get tensor vertex function in $4\times 4$ representation. The full vector and tensor vertex functions in three-dimensions space-time can be expressed in terms of fermion propagators only, which is different from vertex functions in the four-dimensional space-time (in four-dimensions gauge theory, only in the chiral limit, $\Gamma_V^\mu$ and $\Gamma_A^\mu$ at tree level are expressed in terms of the fermion propagators). On the basis of the above, at this point the closedness of the DSE can be established.

\subsection{Two-point Green's Function in 	QED$_3$}
Let us now discuss two important two-point Green's function in 	QED$_3$, namely, the photon propagator and the fermion propagator. The photon propagator can be written as
\begin{flalign}\label{photon}
&iD^{-1}_{\mu \nu}(q) = -q^2[g_{\mu \nu} +(\frac{1}{\lambda} - 1) \frac{q_\mu q_\nu}{q^2}] + \Pi^{\mu \nu}(q),&
 \end{flalign} 
where $q=p_1-p_2$ and $\Pi^{\mu \nu}(q)$ is the photon polarisation vector
\begin{flalign}\label{povec}
&\Pi^{\mu \nu}(q) = \frac{iN_fe^2}{(2\pi)^3}\int d^3p_1 \textit{Tr}_D[\gamma^\mu S_F(p_1)\Gamma^\nu (p_1,p_2)S_F(p_2)],&
\end{flalign} 
where the full fermion propagator  $S^{-1}(p) = \gamma \cdot p A(p^2) + B(p^2)$.
As mentioned above,  the vector vertex function $\Gamma^{\mu}_V$ does not suffer from singularities in the limit $q^2 \to 0$ for $p_1^2 \neq p_2^2$ and $q_\mu \neq 0$ in Minkowski metric. In this case, substituting  the vector vertex function $\Gamma^{\mu}_V$ (\ref{vector later}) into the photon polarisation vector (\ref{povec}), then the photon polarisation vector $\Pi^{\mu \nu}(q) $ is get as showed in Eq. (\ref{povec2}) of the Appendix B.

The DSE for the fermion propagator of QED$_3$ in momentum space,
\begin{flalign}\label{fermdse}
&S_F^{-1}(p_2) =\slashed{p}_2-m - \nonumber \\
&\frac{ie^2}{(2\pi)^3}  \int d^3p_1 \gamma^\mu S(p_1)\Gamma^\nu (p_1,p_2;(p_1-p_2))D_{\nu \mu}(p_1-p_2).&
 \end{flalign} 
 Finally, substituting Eqs. (\ref{photon}-\ref{povec}) into Eq. (\ref{fermdse}), we get the closed DSE for the fermion propagator  in QED$_3$,
\begin{widetext} 
\begin{flalign}\label{fermdse0}    
&2B(p^2_2)=\textit{Tr}S_F^{-1}(p_2) =\textit{Tr}(\slashed{p}_2-m)  - \frac{ie^2}{(2\pi)^3}  \int d^3p_1\textit{Tr}[ \gamma^\mu S(p_1)\Gamma^\nu (p_1,p_2;(p_1-p_2))]D_{\nu \mu}(p_1-p_2).& \nonumber \\
& =-2m  - \frac{ie^2}{(2\pi)^3}  \int d^3p_1\frac{ D_{\nu \mu}(p_1-p_2)}{ ( q^{2}-4m^2)[p^2_1A^2(p^2_1)-B^2(p^2_1)] }\bigg\{ - q^\nu [-2i\epsilon^{\mu \rho \sigma} p_{1\rho} p_{2\sigma}  A(p^2_1)A(p_2^2) +  2p_1^\mu  A(p^2_1)B(p_2^2) &\nonumber \\
& - 2p_2^\mu B(p^2_1)A(p_2^2)] + i   q_{\lambda } \bigg( 2\epsilon^{\mu \nu \lambda}[p_1^2A^2(p^2_1)-B^2(p^2_1)]  +    [i p_{1\rho}p_{2\sigma}M^{\mu \rho \nu \lambda \sigma}(p_1,p_2) A(p^2_1)A(p_2^2) &\nonumber \\
&+4ip_{1\rho}(-g^{\mu \nu}g^{\rho \lambda} + g^{\mu \lambda}g^{\rho \nu })  A(p^2_1)B(p_2^2)   -  i4p_{2 \rho}(g^{\mu \nu}g^{\lambda \rho} -g^{\mu \lambda}g^{\nu \rho}) B(p^2_1)A(p_2^2) - 2\epsilon^{\mu \nu \lambda} B(p^2_1)B(p_2^2) ] \bigg) &\nonumber \\
&+2 m   \bigg( 2g^{\mu \nu}[p_1^2A^2(p^2_1)-B^2(p^2_1)]  +  [4p_{1\rho} p_{2\sigma}(g^{\mu \rho}g^{\nu \sigma} -g^{\mu \nu}g^{\rho \sigma} + g^{\mu \sigma }g^{\rho \nu })   A(p^2_1)A(p_2^2) - 2i\epsilon^{\mu \lambda \nu} p_{1\lambda}    A(p^2_1)B(p_2^2)  &\nonumber \\
&+    2i\epsilon^{\mu \nu \lambda } p_{2\lambda} B(p^2_1)A(p_2^2) - 2g^{\mu \nu} B(p^2_1)B(p_2^2) ]  \bigg)   +   \frac{ [2 m (p_1^\nu+ p_2^\nu)    -i \epsilon^{  \nu \lambda  \rho}  q_\lambda ( p_{1\rho}+p_{2\rho }) ]} {-q_\mu (p_1^\mu+ p_2^\mu  )} \bigg[  q_\tau \bigg( 2g^{\mu \tau}[p^2_1A^2(p_1^2) -B^2(p_1^2)]   &\nonumber \\
& +   [   4 p_{1\rho}p_{2\sigma}(g^{\mu \rho} g^{\tau \sigma} - g^{\mu \tau} g^{ \rho\sigma} +g^{\mu\sigma } g^{ \rho \tau}) A(p_1^2) A(p^2_2)   -2i\epsilon^{\mu \rho \tau} p_{1\rho}  A(p_1^2) B(p_2^2)  +2i\epsilon^{\mu \tau \rho }p_{2\rho} B(p_1^2) A(p^2_2)  - 2g^{\mu  \tau} B(p_1^2)  B(p_2^2) ]\bigg)  &\nonumber \\
&- 2m  [ -2i\epsilon^{\mu \rho \sigma} p_{1\rho}p_{2\sigma} A(p_1^2) A(p^2_2)  + 2p_1^\mu  A(p_1^2) B(p_2^2)  - 2p_2^\mu B(p_1^2) A(p^2_2)  ]  \bigg]&& \nonumber \\
&+i \int \frac{d^3k}{(2\pi)^3}2k_\rho q_\lambda \epsilon^{ \rho \nu \lambda} \frac{1} {-q_\mu (p_1^\mu+ p_2^\mu - 2k^\mu) }\bigg( q_\tau   [ 4 p_{1\rho}p_{3\sigma}(g^{\mu \rho} g^{ \sigma \tau} - g^{\mu \sigma} g^{ \rho \tau} +g^{\mu \tau } g^{ \rho \sigma}) A(p_1^2) A(p^2_3) &\nonumber \\
&+    4p_{1\rho}  p_{4 \sigma}( g^{\mu \rho}g^{\tau \sigma}  -  g^{\mu \tau}g^{\rho \sigma}+ g^{\mu \sigma}g^{ \rho \tau } )  A(p_1^2) A(p^2_4)  -i2 \epsilon^{\mu \rho \tau}  p_{1\rho} A(p_1^2) B(p_3^2) +i2\epsilon^{\mu \rho \tau}p_{3\rho}  B(p_1^2) A(p^2_3)   - 2g^{\mu \tau} B(p_1^2)  B(p_3^2)       &\nonumber \\
&-2i\epsilon^{\mu \rho \tau} p_{1\rho}  A(p_1^2) B(p_4^2) +2i\epsilon^{\mu \tau \rho }p_{4 \rho} B(p_1^2) A(p^2_4)  - 2g^{\mu  \tau} B(p_1^2)  B(p_4^2)  ]   +  2m [ -i2\epsilon^{\mu \rho \sigma} p_{1\rho} p_{3 \sigma}  A(p_1^2) A(p^2_3)  + 2p_1^\mu A(p_1^2) B(p_3^2) &\nonumber \\
&- 2p_3^\mu  B(p_1^2) A(p^2_3)    +2i\epsilon^{\mu \rho \sigma} p_{1\rho}p_{4\sigma} A(p_1^2) A(p^2_4)  - 2p_1^\mu  A(p_1^2) B(p_4^2)  + 2p_4^\mu B(p_1^2) A(p^2_4)    ]  \bigg)\bigg\},&
 \end{flalign} 
and
 \begin{flalign}\label{fermdse1}
&2p_2^\pi A(p_2^2)= \textit{Tr}\gamma^\pi  S_F^{-1}(p_2) = \textit{Tr}(\gamma^\pi \slashed{ p}_2 A(p_2^2) + \gamma^\pi B(p_2^2))  = 2p_2^\pi   - \frac{ie^2}{(2\pi)^3}  \int d^3p_1 \gamma^\pi \gamma^\mu S(p_1)\Gamma^\nu (p_1,p_2;(p_1-p_2))D_{\nu \mu}(p_1-p_2)& \nonumber \\
&= 2p_2^\pi   - \frac{ie^2}{(2\pi)^3}  \int d^3p_1  \frac{ D_{\nu \mu}(p_1-p_2)}{ ( q^{2}-4m^2) [p_1^2A^2(p_1^2)-B^2(p_1^2)]}\bigg\{ q^{\nu }    \bigg[2g^{\pi  \mu }  [p_1^2A^2(p_1^2)-B^2(p_1^2)]-     [4 (g^{ \pi \mu}  p_1  p_2  -p_1^\pi  p_2^\mu  + p_{2}^\pi p_1^\mu ) A(p_1^2) A(p^2_2)  &\nonumber \\
& -2i\epsilon^{\pi \mu \rho }p_{1\rho } A(p_1^2)  B(p_2^2)  +2i\epsilon^{\pi \mu \rho }p_{2\rho }B(p_1^2)A(p^2_2)  -  2g^{\pi \mu} B(p_1^2)B(p_2^2)  ] \bigg]    + i   q_{\lambda } \bigg[4i(-g^{\pi \nu}g^{\mu \lambda} +g^{\pi \lambda}g^{\mu \nu} ) [p_1^2A^2(p_1^2)-B^2(p_1^2)] &\nonumber \\
&+ [i4p_{1\rho}p_{2\sigma}  E^{\pi \mu \rho \nu \lambda \sigma} A(p_1^2) A(p^2_2)  +  ip_{1\rho}Q^{\pi \mu \rho \nu \lambda} A(p_1^2)  B(p_2^2)  - i p_{ 2\rho}M^{\pi \mu  \nu \lambda \rho  } B(p_1^2)A(p^2_2)  - 4i(-g^{\pi \nu }g^{\mu \lambda}+ g^{\pi \lambda}g^{\mu  \nu } )B(p_1^2)B(p_2^2)  ] \bigg] &\nonumber \\
&+2 m   \bigg[-2i\epsilon^{\pi \mu \nu} [p_1^2A^2(p_1^2)-B^2(p_1^2)]   +   [  p_{1\rho}p_{2\sigma}M^{\pi \mu \rho \nu \sigma}A(p_1^2) A(p^2_2)+ 4p_{1\rho} (g^{\pi \mu }g^{\rho \nu} -g^{\pi \rho }g^{\mu \nu} + g^{\pi  \nu}g^{\mu \rho }) A(p_1^2)  B(p_2^2) &\nonumber \\
&-4p_{2 \rho} (g^{\pi \mu }g^{ \nu \rho} -g^{\pi  \nu }g^{\mu \rho} + g^{\pi  \rho}g^{\mu  \nu})B(p_1^2)A(p^2_2)  +2i\epsilon^{\pi \mu \nu}   B(p_1^2)B(p_2^2)  ]        \bigg]   & \nonumber \\
&+\frac{ [2 m (p_1^\nu+ p_2^\nu)    -i \epsilon^{  \nu \lambda  \rho}  q_\lambda ( p_{1\rho}+p_{2\rho }) ] }{-q_\nu (p_1^\nu+ p_2^\nu) } \bigg[ q_\nu     \bigg(-2i\epsilon^{\pi \mu \nu} [p_1^2A^2(p_1^2)-B^2(p_1^2)]   +  [  p_{1\rho}p_{2\sigma}M^{\pi \mu \rho \nu \sigma}A(p_1^2) A(p^2_2) &\nonumber \\
&+ 4p_{1\rho} (g^{\pi \mu }g^{\rho \nu} -g^{\pi \rho }g^{\mu \nu} + g^{\pi  \nu}g^{\mu \rho }) A(p_1^2)  B(p_2^2)  -4p_{2 \rho} (g^{\pi \mu }g^{ \nu \rho} -g^{\pi  \nu }g^{\mu \rho} + g^{\pi  \rho}g^{\mu  \nu})B(p_1^2)A(p^2_2)  +2i\epsilon^{\pi \mu \nu}   B(p_1^2)B(p_2^2)  ]  \bigg) &\nonumber \\
&+2m\bigg(2g^{\pi  \mu} [p_1^2A^2(p_1^2)-B^2(p_1^2)]   -   [4 (g^{ \pi \mu}  p_1  p_2  -p_1^\pi  p_2^\mu  + p_{2}^\pi p_1^\mu ) A(p_1^2) A(p^2_2)   -2i\epsilon^{\pi \mu \rho }p_{1\rho } A(p_1^2)  B(p_2^2) &\nonumber \\
&+2i\epsilon^{\pi \mu \rho }p_{2\rho }B(p_1^2)A(p^2_2)  -  2g^{\pi \mu} B(p_1^2)B(p_2^2)  ]    \bigg) \bigg]  +i \int \frac{d^3k}{(2\pi)^3}\frac{2k_\alpha q_\beta \epsilon^{ \alpha \nu \beta}}{-q_\nu (p_1^\nu+ p_2^\nu-2k^\nu)} \bigg[ q_\nu     \bigg(  [p_{1\rho} p_{3\sigma}M^{\pi \mu \rho \sigma \nu}  A(p_1^2) A(p^2_3) &\nonumber \\
&+ 4p_{1\rho}(g^{\pi \mu}g^{ \rho \nu} -g^{\pi \rho}g^{ \mu  \nu} +g^{\pi \nu}g^{ \mu \rho}  )   A(p_1^2)  B(p_3^2)  -  4p_{3\rho}(g^{\pi \mu}g^{ \rho \nu} -g^{\pi \rho}g^{ \mu  \nu} +g^{\pi \nu}g^{ \mu \rho}  )  B(p_1^2) A(p^2_3)+i2\epsilon^{\pi \mu \nu}  B(p_1^2)B(p_3^2)  ] &\nonumber \\
&+ [  p_{1\rho}p_{4\sigma}M^{\pi \mu \rho \nu \sigma}A(p_1^2) A(p^2_4)  + 4p_{1\rho} (g^{\pi \mu }g^{\rho \nu} -g^{\pi \rho }g^{\mu \nu} + g^{\pi  \nu}g^{\mu \rho }) A(p_1^2)  B(p_4^2)  -4p_{2 \rho} (g^{\pi \mu }g^{ \nu \rho} -g^{\pi  \nu }g^{\mu \rho} + g^{\pi  \rho}g^{\mu  \nu})B(p_1^2)A(p^2_4)  & \nonumber \\
&+2i\epsilon{\pi \mu \nu}   B(p_1^2)B(p_4^2)  ]  \bigg)  +2m\bigg( [4 (g^{ \pi \mu}  p_1  p_3  -p_1^\pi  p_3^\mu  + p_{3}^\pi p_1^\mu ) A(p_1^2) A(p^2_3)   -2i\epsilon^{\pi \mu \rho }p_{1\rho } A(p_1^2)  B(p_3^2)  +2i\epsilon^{\pi \mu \rho }p_{3 \rho }B(p_1^2)A(p^2_3)  &\nonumber \\
&-  2g^{\pi \mu} B(p_1^2)B(p_3^2)  ]  -   [4 (g^{ \pi \mu}  p_1  p_4  -p_1^\pi  p_4^\mu  + p_{4}^\pi p_1^\mu ) A(p_1^2) A(p^2_4)    -2i\epsilon^{\pi \mu \rho }p_{1\rho } A(p_1^2)  B(p_4^2)  &\nonumber \\
&+2i\epsilon^{\pi \mu \rho }p_{4\rho }B(p_1^2)A(p^2_4)  -  2g^{\pi \mu} B(p_1^2)B(p_4^2)  ]    \bigg) \bigg] \bigg\},& 
 \end{flalign}
where
 \begin{flalign}   
&  E^{\pi \mu \rho \nu \lambda \sigma}= \frac{1}{8}[\textit{Tr} ( \gamma^\pi \gamma^\mu \gamma^\rho \gamma^\nu \gamma^\lambda  \gamma^\sigma ) - \textit{Tr} ( \gamma^\pi \gamma^\mu \gamma^\rho \gamma^\lambda  \gamma^\nu \gamma^\sigma )]  & \nonumber \\
 &= g^{\pi \mu}(g^{\rho \nu}g^{\lambda \sigma} - g^{\rho\lambda}g^{  \nu\sigma})  -   g^{\pi \rho} (g^{\mu \nu}g^{\lambda \sigma} - g^{\mu \lambda}g^{\nu \sigma} )   + g^{\pi \nu} (g^{\mu \rho} g^{ \lambda \sigma}   - g^{\mu \lambda} g^{ \rho \sigma} + g^{\mu \sigma } g^{ \rho \lambda} )  -g^{\pi \lambda} (g^{\mu \rho} g^{ \nu \sigma} - g^{\mu \nu } g^{ \rho \sigma} + g^{\mu \sigma} g^{\rho \nu })  &\nonumber \\
 & + g^{\pi \sigma} (-g^{\mu \nu}g^{\rho \lambda} + g^{\mu \lambda}g^{\rho \nu} ) , & \nonumber \\
& Q^{\pi \mu \rho \nu \lambda} = \frac{1}{2}[\textit{Tr} ( \gamma^\pi \gamma^\mu \gamma^\rho  \gamma^\nu \gamma^\lambda )  -\textit{Tr} ( \gamma^\pi \gamma^\mu \gamma^\rho  \gamma^\lambda \gamma^\nu ) ] =  \frac{1}{2}(M^{\pi \mu \rho \nu \lambda} -M^{\pi \mu  \rho  \lambda \nu}) & \nonumber \\
&=-2ig^{ \pi \mu}\epsilon^{\rho \nu \lambda}  + 2ig^{ \pi \rho}\epsilon^{ \mu\nu \lambda}  - ig^{ \pi \nu}\epsilon^{\mu \rho  \lambda} +i2\epsilon^{\pi \lambda m} (g^{\mu \rho} g^\nu_m - g^{\mu \nu} g^\rho_m+ g^{\mu}_m g^{\rho \nu})   +  ig^{ \pi  \lambda}\epsilon^{\mu \rho \nu} - i2\epsilon^{\pi \nu m} (g^{\mu \rho} g^\lambda_m - g^{\mu \lambda} g^\rho_m+ g^{\mu}_m g^{\rho \lambda}) , & \nonumber \\
& \frac{1}{2}(M^{\pi \mu \rho \nu \lambda} -M^{\pi \mu \nu   \rho  \lambda}) =M^{\pi \mu \rho \nu \lambda}. &
\end{flalign}
\end{widetext}
 Based on the coupled equations satisfied by $A(p^2)$ and $B(p^2)$ above, in principle we can strictly solve the complete fermion propagator by numerical iterative method, then the chiral symmetry spontaneous breaking and confinement characteristics of QED$_3$ can be analytically analyzed. But what needs to be pointed out here is that the coupled integral equations Eqs. (\ref{fermdse0}, \ref{fermdse1}) are extremely complex, which is a huge challenge for the rigorous numerical solutions. We will work on this problem in future work.

\section{Summary and Conclusion}

To summarize, we first derive the transverse WTI of \textit{N}-dimensional gauge theory by means of the canonical quantization method and the path 
integration method, and then using the characteristics of the $\gamma$ matrix representation in three-dimensional gauge theory it is shown that the normal (longitudinal) WTI together with the transverse WTI form a complete set of Ward-Takahashi type constraint relations for the fermion-boson vertex functions in QED$_3$ theory. By solving this complete set, the full scalar, vector and tensor vertex functions ($\Gamma_S, \Gamma^\mu_V, \Gamma_{T}^{\mu \nu}$) can be expressed in terms of the fermion's two-point functions, which is completely different from the situation in four-dimensions gauge theory (where only in the chiral limit, $\Gamma_V^\mu$ and $\Gamma_A^\mu$ at tree level are expressed in terms of the fermion propagators). It is found that the full tensor vertex function in $4\times 4$ representation is different from that in $2\times 2$ representation. This means that when we study the dynamic behavior of three-dimensional gauge theory related to the tensor vertex function, we must specify the $\gamma$ matrix representation in advance. Furthermore, we examine the possible kinematic singularities that the dressed vertex function may have, and find that the vector vertex function $\Gamma^{\mu}_V$  does not suffer from singularities in the limit $q^2 \to 0$ for $p_1^2 \neq p_2^2, q_\mu \neq 0$ in Minkowski metric. Then substituting  the vector vertex function into the DSEs for the fermion propagator and photon propagator, finally we get the closed DSE in QED$_3$. Based on this set of closed coupled nonlinear integral equations, in principle we can numerically solve the two-point Green functions and three-point Green functions by  numerical iteration method to analyze the mechanism of the chiral symmetry spontaneous breaking and confinement in QED$_3$.

Finally, we need to emphasize that low-dimensional gauge theory has a very wide range of applications in condensed matter physics. In particular, QED$_3$ has been suggested to be the effective low-energy field theory for the anomalous normal state of high-$T_c$ cuprate superconductors \cite{wrantner01, PhysRevLett87257003, PhysRevLett88047006}. It also provides a promising field-theoretic description for such exotic quantum many-body state as $U(1)$ quantum spin liquid \cite{spinliquid}. When massless Dirac fermions are coupled to $U(1)$ gauge boson, they acquire a finite anomalous dimension due to the strong gauge interaction \cite{wrantner01, PhysRevLett87257003, PhysRevLett88047006, wrantner}. This may lead
to intriguing Luttinger-like behaviors, which has been used to understand the absence of well-defined quasiparticle peaks in the normal state of high-$T_c$ cuprate superconductors \cite{wrantner01,  PhysRevLett87257003, PhysRevLett88047006, wrantner}. To reveal the nature of these Luttinger-like behaviors, one needs to compute certain types of Green's function very carefully. The gauge invariance must be preserved during the analytical calculations \cite{khvecomment, khveprb02, khvenpb02, gusunin03, franz03}. In principle, these Green's functions can be self-consistently obtained by solving a close set of DSEs. We expect that the generic WTI obtained in this work would be utilized to calculate the gauge invariant Green's functions by means of DSEs.

\begin{acknowledgments}
We thank Prof. Guo-Zhu Liu for very helpful discussions.  This work is supported in part by the National Natural Science Foundation of China (11475085, 11535005, 11690030), the National Major state Basic Research and Development (2016YFE0129300) and the Anhui Provincial Natural Science Foundation (1908085MA15).
\end{acknowledgments}

\appendix

\section{}
First one introduces two bilinear covariant current operators, 
\begin{flalign} \label{bilinear}
&V^{\rho \mu \nu \lambda} (x)=\frac{1}{4} \bar{\psi}(x) \bigg[[\gamma^\rho, \sigma^{\mu \nu}], \gamma^\lambda  \bigg] \psi(x) =g^{\rho \mu} j^{\nu \lambda}(x) -g^{\rho \nu} j^{\mu \lambda}(x), & \nonumber  \\
&V^{\rho \mu \nu} (x) =\frac{-i}{2} \bar{\psi}  [\gamma^\rho, \sigma^{\mu \nu}] \psi =g^{\rho \mu} j^\nu(x) -g^{\rho \nu} j^\mu(x) .& 
\end{flalign}
In the canonical quantization method, one notes here the general identity \cite{proof} 
\begin{flalign}\label{zhengzhe}
&\partial^x_\lambda \langle 0|TV^{\lambda \mu \nu (\alpha)} (x)  \psi(x_1)\bar{\psi}(y_1)\dots\psi(x_n)\bar{\psi}(y_n) |0\rangle &\nonumber \\
&=\sum_{i=1}^n  \delta_{\lambda 0}\langle 0|T\bigg\{ V^{\lambda \mu \nu (\alpha ) } (x) ,\psi(x_i)] \delta(x^0-x_i^0)\bar{\psi}(y_i) &\nonumber \\
&+\psi(x_i)[V^{\lambda \mu \nu (\alpha ) } (x),\bar{\psi}(y_i)] \delta(x^0-y_i^0) \bigg\}& \nonumber \\
&\times \psi(x_1)\bar{\psi}(y_1)\dots  \ulcorner \psi(x_i)\bar{\psi}(y_i) \lrcorner\dots \psi(x_n)\bar{\psi}(y_n) |0\rangle &\nonumber  \\
& + \langle 0| T\partial^x_\lambda V^{\lambda \mu \nu (\alpha ) } (x) \psi(x_1)\bar{\psi}(y_1)\dots\psi(x_n)\bar{\psi}(y_n)  |0\rangle, &
\end{flalign}
where the delimiter $ \ulcorner \; \lrcorner $ term above  means its omission.  The last term in above equation leads to a similar situation of $\langle 0| T\bar{\psi}(x) N (\overrightarrow{\partial}_\lambda^x +\overleftarrow{\partial}_\lambda^x)\psi(x)  \psi(y)\bar{\psi}(z)  |0\rangle $, normally, where $N$ is matrix with an anti-communication relation. It means that the transverse WT identity exhibits different appearance depending on the dimensionality of space-time, because the anti-communication relation depends on the space-time dimension.

Substituting the relations (\ref{bilinear}) into Eqs. (\ref{zhengzhe}), there are
\begin{flalign}\label{sl1}
&\partial_\rho \langle 0| T V^{ \rho\mu \nu}(x) \psi(y) \bar{\psi}(z)|0\rangle &\nonumber \\
&=  \partial^\mu   \langle 0| T j^\nu (x)  \psi(y) \bar{\psi}(z)|0\rangle  - \partial^\nu   \langle 0| T j^\mu (x) \psi(y) \bar{\psi}(z)|0\rangle  &\nonumber \\
& =-\delta^4(x-y) \gamma^0 \frac{i}{2} [\sigma^{\mu \nu},\gamma^0] \langle 0| T \psi(x)  \bar{\psi}(z)  |0\rangle & \nonumber \\
&+  \langle 0  |T\psi(y)  \bar{\psi}(x) | 0\rangle \frac{i}{2} [\sigma^{\mu \nu}, \gamma^0]\gamma^0 \delta^3(x-z) &\nonumber \\
&+ \langle 0| T  \partial_\rho  V^{ \rho\mu \nu}(x) \psi(y) \bar{\psi}(z)|0\rangle &
\end{flalign}
and
\begin{flalign}  \label{b3}
&\partial_\rho \langle 0| T V^{ \rho\mu \nu \lambda}(x) \psi(y) \bar{\psi}(z)|0\rangle &\nonumber \\
& = \partial^\mu   \langle 0| T j^{\nu \lambda}(x)  \psi(y) \bar{\psi}(z)|0\rangle  - \partial^\nu   \langle 0| T j^{\mu \lambda} (x) \psi(y) \bar{\psi}(z)|0\rangle & \nonumber \\
&=  - \gamma^0 \frac{1}{4} \bigg[[ \gamma^0, \sigma^{\mu \nu}], \gamma^\lambda\bigg] \delta^4(x-y)   \langle 0| T\psi(x) \bar{\psi}(z) |0\rangle &\nonumber \\
&+\langle 0|T  \psi(y) \bar{\psi}(x) | 0\rangle \frac{1}{4}\bigg[ [\gamma^0, \sigma^{\mu \nu}],\gamma^\lambda \bigg] \gamma^0 \delta^4(x-z) &\nonumber \\
&+\langle 0| T \partial_\rho V^{\rho \mu \nu \lambda}(x) \psi(y) \bar{\psi}(z) | 0\rangle.&
\end{flalign}

In order to relate the last term in the above equation to a definite Green's function and to make the equations above more concise, here one needs to consider  two conditions. Firstly, the equation of motion for fermions with mass $\bar{\psi}(i\overleftarrow{\slashed{D}} + m) = 0, (i\overrightarrow{\slashed{D}} - m)\psi = 0$ are introduced to make the last term more concise. So the term $\gamma^\mu \partial_\mu \psi(x) $ and  $\partial_\mu  \bar{\psi}(x) \gamma^\mu$ need to be shown in the equations as
\begin{flalign} \label{v2d}
&\langle 0| T \partial_\rho V^{\rho \mu \nu}(x) \psi(y) \bar{\psi}(z) | 0\rangle &\nonumber \\
&= \langle 0|  i\bar{\psi}(x)  \sigma^{\mu \nu} \gamma^\rho  \partial_\rho \psi(x) \psi(y) \bar{\psi}(z)|0\rangle& \nonumber \\
&-  \langle 0| i \partial_\rho \bar{\psi}(x) \gamma^\rho   \sigma^{\mu \nu} \psi(x) \psi(y) \bar{\psi}(z)| 0\rangle  & \nonumber \\
&+ \langle 0|  \bar{\psi}(x) \frac{i}{2} \{ \sigma^{\mu \nu} ,\gamma^\rho \}(\overleftarrow{\partial}_\rho -\overrightarrow{\partial}_\rho) \psi(x)  \psi(y) \bar{\psi}(z) | 0\rangle      &
\end{flalign}
and
\begin{flalign} \label{tsd}
&\langle 0| T \partial_\rho V^{\rho \mu \nu \lambda}(x) \psi(y) \bar{\psi}(z) | 0\rangle &\nonumber \\
& =\frac{1}{4} \langle 0|  \partial_\rho[\bar{\psi}(x)(  \gamma^\rho \sigma^{\mu \nu}\gamma^\lambda + \sigma^{\mu \nu}\gamma^\lambda\gamma^\rho )  \psi(x) ] \psi(y)\psi(z) |0\rangle & \nonumber \\
&  +   \frac{1}{4} \langle 0|  \partial_\rho [ \bar{\psi}(x) (  \gamma^\rho  \gamma^\lambda\sigma^{\mu \nu} +  \gamma^\lambda\sigma^{\mu \nu} \gamma^\rho) \psi(x) ]\psi(y) \bar{\psi}(z) |0\rangle & \nonumber \\
&-\langle 0|  \partial_\rho \bigg\{ \bar{\psi}(x)g^{\rho \lambda}\sigma^{\mu \nu}   \psi(x) \bigg\} \psi(y)\psi(z) |0\rangle.&
\end{flalign}
To further simplify the calculations, here one needs to use the following relations to the first item of Eq. (\ref{tsd}). 
 \begin{flalign}\label{f1}
&\frac{1}{4} \langle 0|  \partial_\rho [\bar{\psi}(x)(  \gamma^\rho \sigma^{\mu \nu}\gamma^\lambda + \sigma^{\mu \nu}\gamma^\lambda\gamma^\rho )  \psi(x)] \psi(y)\psi(z) |0\rangle & \nonumber \\
&=\frac{1}{2} \langle 0|  \partial_\rho  \bar{\psi}(x) \gamma^\rho \sigma^{\mu \nu}\gamma^\lambda  \psi(x)   \psi(y)\psi(z) |0\rangle &\nonumber \\
 &+  \frac{1}{2} \langle 0|   \bar{\psi}(x) \sigma^{\mu \nu}\gamma^\lambda   \gamma^\rho \partial_\rho\psi(x)   \psi(y)\psi(z) |0\rangle &\nonumber \\
&- \langle 0|  \bar{\psi}(x) A^{\rho \mu \nu \lambda} ( \overleftarrow{\partial^x_\rho} -  \overrightarrow{\partial^x_\rho} ) \psi(x)   \psi(y)\psi(z) |0\rangle,  &
   \end{flalign}
where one has defined $\frac{1}{4} [ \gamma^\rho, \sigma^{\mu \nu}\gamma^\lambda] =A^{\rho \mu \nu \lambda}$. With the similar procedure one derives the second item of Eq. (\ref{tsd}), and defined $\frac{1}{4} [ \gamma^\rho, \gamma^\lambda \sigma^{\mu \nu}] =B^{\rho \lambda \mu \nu }$.

Secondly,  one needs to move the derivative operators out of the $T$-product. For this purpose, one can write the form $\langle 0|T \bar{\psi}(x)N \psi(x)  \psi(y)\bar{\psi}(z)  |0\rangle $ as $\langle 0|T \bar{\psi}(x')N \psi(x)  \psi(y)\bar{\psi}(z)  |0\rangle $ and then take $x'\to x$. The above new expression including the nonlocal current is not gauge invariant. It needs to introduce a Wilson line $U(x,x') = P\mathrm{exp}[-ig\int^{x'}_x dy^\rho A_\rho (y)]$, joining the two space-time points ($x,x'$) to ensure that the current operators are locally gauge invariant. Comprehensive use of the Wilson line,  the Eq. (\ref{zhengzhe}) and the equation of motion for fermions, there eventually are two relations
  \begin{flalign}\label{gs1}
& (\partial_\rho^{x'} + \partial_\rho^x)\langle 0|  T\bar{\psi}(x')M^{\rho \mu \nu \lambda  } U(x',x)\psi(x)  \psi(y) \bar{\psi}(z) | 0\rangle &\nonumber \\
& =\langle 0|  T \bar{\psi}(x)  M^{ \rho \mu \nu \lambda }(\overleftarrow{\partial_\rho^x} +\overrightarrow{\partial_{\rho}^x}) \psi(x)  \psi(y) \bar{\psi}(z) | 0\rangle & \nonumber \\
&-\delta^4(x -y) \gamma^0 M^{0 \mu \nu \lambda}  \langle 0|  T \psi(x)  \bar{\psi}(z) | 0\rangle \nonumber \\
& + \langle 0|  T\psi(y)  \bar{\psi}(x)    |0 \rangle M^{0 \mu \nu \lambda}  \gamma^0\delta^4(x-z)  &
\end{flalign}
and 
  \begin{flalign}\label{gs2}
& (\partial_\rho^{x'} - \partial_\rho^x)\langle 0|  T\bar{\psi}(x')M^{\rho \mu \nu \lambda  } U(x',x)\psi(x)  \psi(y) \bar{\psi}(z) | 0\rangle & \nonumber \\
 &=\langle 0|  T \bar{\psi}(x)  M^{ \rho \mu \nu \lambda }(\overleftarrow{\partial_\rho^x} -\overrightarrow{\partial_{\rho}^x}) \psi(x)  \psi(y) \bar{\psi}(z) | 0\rangle &\nonumber \\
&-\delta^4(x -y) \gamma^0 M^{0 \mu \nu \lambda}  \langle 0|  T \psi(x)  \bar{\psi}(z) | 0\rangle &\nonumber \\
& - \langle 0|  T\psi(y)  \bar{\psi}(x)    |\rangle M^{0 \mu \nu \lambda}  \gamma^0\delta^4(x-z) 0 & \nonumber \\
&-2igA_\rho \langle 0|  T \bar{\psi}(x) M^{\rho \mu \nu \lambda} \psi(x)  \psi(y) \bar{\psi}(z) | 0\rangle, &
\end{flalign}
where $ M^{ \rho \mu \nu \lambda }$ denotes a matrix.

Taking into account the above equations, substituting relations (\ref{f1},\ref{gs1},\ref{gs2}) into relations (\ref{sl1}, \ref{b3}, \ref{v2d}, \ref{tsd}) we arrive at the transverse WT relations for the fermion’s vertex functions in gauge theories in  configuration space
\begin{flalign} \label{vector current}
& \partial^\mu   \langle 0| T j^\nu (x)  \psi(y) \bar{\psi}(z)|0\rangle  - \partial^\nu   \langle 0| T j^\mu (x) \psi(y) \bar{\psi}(z)|0\rangle &\nonumber \\
&= \lim_{x'\to x}( \partial^{x'}_\rho - \partial^{x}_\rho)\langle 0|  T \bar{\psi}(x')\frac{i}{2}\{\gamma^\rho, \sigma^{\mu \nu}\}  U(x',x)\psi(x) \psi(y)\bar{\psi}(z) |0\rangle &\nonumber \\ 
&+i \sigma^{\mu \nu}  \delta^4(x-y) \langle 0|   T \psi(x)  \bar{\psi}(z)  |0\rangle  &\nonumber \\
&+ i\langle 0|  T\psi(y)  \bar{\psi}(x)   0\rangle  \sigma^{\mu \nu }  \delta^4(x-z) & \nonumber \\
&+2m   \langle 0| T \bar{\psi}(x)   \sigma^{\mu \nu} \psi(x)  \psi(y)\bar{\psi}(z) |0\rangle &  
\end{flalign}
and
\begin{flalign} \label{tensor current}
& \partial^\mu   \langle 0| T j^{\nu \lambda}(x)  \psi(y) \bar{\psi}(z)|0\rangle  - \partial^\nu   \langle 0| T j^{\mu \lambda} (x) \psi(y) \bar{\psi}(z)|0\rangle& \nonumber \\
&= -\frac{1}{2}\{\sigma^{\mu \nu} , \gamma^\lambda\} \delta^4(x-y)   \langle 0| T\psi(x) \bar{\psi}(z) |0\rangle & \nonumber \\
&+\langle 0|T  \psi(y) \bar{\psi}(x) | 0\rangle \frac{1}{2}\{\sigma^{\mu \nu} , \gamma^\lambda\} \delta^4(x-z) &\nonumber \\
&-  (\partial_\rho^{x'}-\partial_\rho^{x})  \langle 0|  T\bar{\psi}(x') \frac{1}{4}\bigg[  \gamma^\rho, \bigg\{ \sigma^{\mu \nu}, \gamma^\lambda  \bigg\}\bigg] U(x',x)\psi(x)  \psi(y) \bar{\psi}(z) | 0\rangle & \nonumber \\
 &-(\partial^{\lambda(x')}+ \partial^{\lambda(x)})\langle 0|  T\bar{\psi}(x')\sigma^{\mu \nu  } U(x',x)\psi(x)  \psi(y) \bar{\psi}(z) | 0\rangle. &
\end{flalign}

In the path integration method \cite{tward}, in Abelian case there are the identity
\begin{flalign}\label{path}
&\langle i\gamma^\mu [\partial_\mu -ieA_\mu(x)]\psi(x) - m\psi(x) +\eta(x) \rangle_J =0& \nonumber \\
&\langle \bar{\psi}(x) i\gamma^\mu [\overleftarrow{\partial_\mu} + ieA_\mu(x)] + m\bar{\psi}(x) -\bar{\eta}(x) \rangle_J =0. &
\end{flalign}
So one needs only to pay attention to the fermionic part
\begin{flalign}
&\mathcal{L}_F=\bar{\psi}i\gamma^\mu(\partial_\mu -ieA_\mu)\psi -\bar{\psi}m\psi + \bar{\eta}\psi +\bar{\psi}\eta. &
\end{flalign}
If one identifies $A_\mu$ in $\mathcal{L}_F$ as  $A_\mu =A_\mu^\alpha T^\alpha$ with the generator $T^\alpha$ of the gauge group $G$, the following relations also hold for the non-Abelian case irrespective of the gauge part. Then one can multiply Eq. (\ref{path}) by the
matrix $S$ from the left  (right), where $S$ may be a matrix of spinor, flavors and colors spaces. Operating the differential operator $\frac{\delta}{\delta \eta(y)}(\frac{\delta}{\delta \bar{\eta}(y)})$ to the resulting equation,  an then plus or minus, subsequently taking derivatives of both side with respect to $\frac{\delta}{\delta \bar{\eta} (y)}$ and $\frac{\delta}{\delta  \eta (z)}$ and setting all the source terms to zero, one gets the transverse WT identity
\begin{flalign} \label{zonggongshi1}
 &\partial_\rho \langle \bar{\psi}(x)\frac{i}{2}\{S, \gamma^\rho \}\psi(x);  \psi(y)\bar{\psi}(z) \rangle_c & \nonumber \\
 &= -   \langle \bar{\psi}(x)\frac{i}{2}[S, \gamma^\rho ] (\overrightarrow{\partial}_\rho - \overleftarrow{\partial}_\rho)\psi(x); \psi(y)\bar{\psi}(z) \rangle_c  & \nonumber \\
&-e \langle \bar{\psi}(x)[ S, \gamma^\rho A_\rho]\psi(x);  \psi(y)\bar{\psi}(z) \rangle_c  & \nonumber \\
&+  \langle \bar{\psi}(x)[S, m] \psi(x); \psi(y)\bar{\psi}(z)  \rangle_c & \nonumber \\
& +  \langle \psi(y)\bar{\psi} (x)   \rangle_c  S \delta^d(x-z)  +   S\langle \psi(x)  \bar{\psi}(z)  \rangle_c\delta^d(x-y)  & 
\end{flalign}
and
\begin{flalign}\label{zonggongshi2}
& \partial_\rho \langle \bar{\psi}(x)\frac{i}{2} [S, \gamma^\rho] \psi(x); \psi(y) \bar{\psi}(z)\rangle_c & \nonumber \\
&=  - \langle \bar{\psi}(x)\frac{i}{2}\{S, \gamma^\rho \} (\overrightarrow{\partial}_\rho - \overleftarrow{\partial}_\rho)\psi (x); \psi(y)  \bar{\psi}(z) \rangle_c  &\nonumber \\
&-e \langle \bar{\psi}(x) \{S, \gamma^\rho A_\rho \}\psi(x);  \psi(y)  \bar{\psi}(z) \rangle_c &\nonumber \\
& +  \langle \bar{\psi}(x)\{ S, m\} \psi(x);  \psi(y)  \bar{\psi}(z)\rangle_c & \nonumber \\
&  -  \langle  \psi(y) \bar{\psi}(x)   \rangle_c  S \delta^d(x-z)  - S  \langle  \psi(x) \bar{\psi}(z) \rangle_c\delta^d(x-y).  &
\end{flalign}
Let $S =S_s \otimes  S_f \otimes S_c $ be a direct product of operators within the space of spinor, flavor and color.  If choose $S =I_s \otimes I_f \otimes I_c $, one obtains the normal WT identity. It turns out that the transverse WT identity for vector current is obtained from Eq. (\ref{zonggongshi1}) by  choosing $ S =\sigma_{\mu \nu} \otimes I_f \otimes I_c $. 

From the derivation of the above formula, the transverse WT identity exhibits different appearance depending on the dimensionality of space–time. However, it is not easy to calculate the transverse WT identity for tensor current. In order to use the above relations (\ref{zonggongshi1}, \ref{zonggongshi2}), we need to modify the bilinear covariant current operators (\ref{bilinear}) slightly:
\begin{flalign} \label{bilinear2}
&V^{\rho \mu \nu \lambda} (x)=\frac{1}{4} \bar{\psi}(x) \bigg[[\gamma^\rho, \sigma^{\mu \nu}], \gamma^\lambda  \bigg] \psi(x) =g^{\rho \mu} j^{\nu \lambda}(x) -g^{\rho \nu} j^{\mu \lambda}(x) &   \nonumber \\
 &=\bar{\psi}(x)\frac{1}{4} \bigg\{\gamma^\rho,  \{\sigma^{\mu \nu}, \gamma^\lambda\} \bigg\}\psi(x) -\bar{\psi}(x)g^{\rho \lambda} \sigma^{\mu \nu }\psi(x).&  
\end{flalign}
Through the above relations ( \ref{zonggongshi1}, \ref{bilinear2}), the transverse WTI for fermion's vertex functions can be obtained by
\begin{flalign}
&\partial_\rho^x \langle 0| TV^{\rho \mu \nu \lambda} (x)\psi(y) \bar{\psi}(z) |0\rangle &\nonumber \\
& =  \partial_\rho \langle \bar{\psi}(x)\frac{i}{2}\{S, \gamma^\rho \}\psi(x);  \psi(y)\bar{\psi}(z) \rangle_c &\nonumber \\
&   -\partial^\lambda \langle \bar{\psi}(x)  \sigma^{\mu \nu } \psi(x);  \psi(y)\bar{\psi}(z) \rangle_c  & \nonumber \\
&=\partial_x^\mu \langle 0|Tj^{\nu \lambda} (x)\psi(y) \bar{\psi}(z)|0\rangle  - \partial_x^\nu \langle 0|Tj^{\mu \lambda}  (x) \psi(y) \bar{\psi}(z)|0\rangle,& 
\end{flalign}
where $S = \frac{-i}{2}\{\sigma^{\mu \nu }, \gamma^\lambda \}$. Then it can be verified that the the transverse WT identity (\ref{vector current}, \ref{tensor current}) are obtained by the path integration method (\ref{zonggongshi1}, \ref{zonggongshi2}). 

As shown above, the transverse and longitudinal WT identities in the four-dimensional gauge theory do not specify the vertex function with a two-point Green`s function, thus forming a closed DSEs.
But in the case of low-dimension gauge theory,
such as QED$_3$, the basic situation has changed a lot.
In QED$_3$ theory, one can find a sets of transverse WT relations (for the vector and tensor vertex function) are coupled to each other, the transverse relations together with the longitudinal WT identities would lead to a complete set of WT-type constraint relations for the three-point functions. Then the complete expressions for three vertex functions can be deduced by solving this complete set of WT relations. 

\section{}
As mentioned above,  substituting  the vector vertex function $\Gamma^{\mu}_V$ (\ref{vector later}) into the photon polarisation vector (\ref{povec}), then the photon polarisation vector is get as follows
\begin{widetext} 
\begin{flalign}\label{povec2}
&\Pi^{\mu \nu}(q) = \frac{iN_fe^2}{(2\pi)^3}\int d^3p_1 \textit{Tr}_D[\gamma^\mu S_F(p_1)\Gamma^\nu (p_1,p_2)S_F(p_2)].&\nonumber \\
& = \frac{iN_fe^2}{(2\pi)^3}\int d^3p_1 \frac{1}{ ( q^{2}-4m^2) }\textit{Tr} \bigg\{ q^{\nu }  [  \frac{2p_2^\mu  A(p_2^2) }{p_2^2A^2(p_2^2)-B^2(p_2^2)}  -    \frac{2p_1^\mu  A(p_1^2) }{p_1^2A^2(p_1^2)-B^2(p_1^2)} ]  & \nonumber \\
  &+ i   q_{\lambda } \bigg(\frac{i 4p_{2\rho}[g^{\mu  \nu}g^{\lambda \rho} - g^{\mu \lambda}g^{ \nu \rho}] A(p_2^2)-  2\epsilon^{\mu \nu \lambda}B(p_2^2)}{p_2^2A^2(p_2^2)-B^2(p_2^2)}+ \frac{i4p_{1\rho}[- g^{\mu  \nu}g^{ \rho \lambda}+ g^{\mu \lambda}g^{\rho \nu }] A(p_1^2)- 2\epsilon^{\mu \nu \lambda}B(p_1^2)}{p_1^2A^2(p_1^2)-B^2(p_1^2)} \bigg) &\nonumber \\
  &+2 m  [ \frac{ -i2p_{2 \rho} \epsilon^{\mu \nu \rho}A(p_2^2)- 2g^{\mu \nu} B(p_2^2)}{p_2^2A^2(p_2^2)-B^2(p_2^2)}  + \frac{-i2\epsilon^{\mu \rho \nu} p_{1\rho} A(p_1^2) - 2g^{\mu  \nu} B(p_1^2) }{p_1^2A^2(p_1^2)-B^2(p_1^2)} ] &\nonumber \\
& + \frac{[2 m (p_1^\nu+ p_2^\nu)    -i \epsilon^{ \rho \nu \lambda}  q_\lambda ( p_{1\rho}+p_{2\rho }) ]}{-q_\tau (p_1^\tau+ p_2^\tau) }\bigg( q_\tau  [\frac{ -i2p_{2 \rho} \epsilon^{\mu \tau \rho}A(p_2^2)- 2g^{\mu \tau} B(p_2^2)}{p_2^2A^2(p_2^2)-B^2(p_2^2)}   +\frac{-i2\epsilon^{\mu \rho \tau} p_{1\rho} A(p_1^2) - 2g^{\mu  \tau} B(p_1^2) }{p_1^2A^2(p_1^2)-B^2(p_1^2)}   ] &\nonumber \\
&+ 2m [ \frac{2p_2^\mu  A(p_2^2) }{p_2^2A^2(p_2^2)-B^2(p_2^2)}  - \frac{2p_1^\mu  A(p_1^2) }{p_1^2A^2(p_1^2)-B^2(p_1^2)} ] \bigg) & \nonumber \\
&+i \int \frac{d^3k}{(2\pi)^3}2k_\alpha q_\beta \epsilon^{ \alpha \nu \beta } \frac{q_\tau [C^{\mu \tau}(p_3,p_1,p_2) + O^{\mu \tau}(p_4,p_1,p_2)]+2m [F^\mu(p_3,p_1,p_2) -F^\mu(p_4,p_1,p_2) }{-q_\sigma (p_3^\sigma+ p_4^\sigma )  }\bigg\},   &
\end{flalign} 
where we have used this relationship $q=p_1-p_2,   p_3 =p_1-k,   p_4=p_2-k$, and the relations of $C^{\mu \tau}(p_3), O^{\mu \tau}(p_4),  F^\mu( p_3),  M^{\mu \nu \rho \tau \lambda} $ are defined as follows:
\begin{flalign}
&C^{\mu \tau}(p_3,p_1,p_2)=\textit{Tr} \gamma^\mu S_F(p_1) S^{-1}_F(p_1-k) \gamma^\tau S_F(p_2) & \nonumber \\
&=  \frac{1}{[p_1^2A^2(p_1^2)-B^2(p_1^2)][p_2^2A^2(p_2^2)-B^2(p_2^2)]} \times &\nonumber \\
&  [ p_{1\rho} p_{3\sigma}  p_{2\lambda}M^{\mu \rho \sigma \tau \lambda} A(p_1^2) A(p_2^2)A(p_3^2) +4 p_{1\rho}p_{2\sigma}(g^{\mu \rho} g^{\tau \sigma} - g^{\mu \tau} g^{ \rho\sigma} +g^{\mu\sigma } g^{ \rho \tau})A(p_1^2) A(p_2^2) B(p_3^2)  \nonumber \\
&- 4 p_{3\rho}p_{2\sigma}(g^{\mu \rho} g^{\tau \sigma} - g^{\mu \tau} g^{ \rho\sigma} +g^{\mu\sigma } g^{ \rho \tau}) B(p_1^2)A(p_2^2)A(p_3^2)  +2i\epsilon^{\mu \tau \rho} p_{2\rho}   B(p_1^2)A(p_2^2)B(p_3^2)&\nonumber \\
&-4 p_{1\rho}p_{3\sigma}(g^{\mu \rho} g^{ \sigma \tau} - g^{\mu \sigma} g^{ \rho \tau} +g^{\mu \tau } g^{ \rho \sigma}) A(p_1^2)B(p_2^2) A(p_3^2)  +2i\epsilon^{\mu \rho \tau } p_{1\rho} A(p_1^2)  B(p_2^2) B(p_3^2)  -2i\epsilon^{\mu \rho \tau } p_{1\rho} B(p_1^2)B(p_2^2)A(p_3^2)  &\nonumber \\
&+2g^{\mu \tau} B(p_1^2)B(p_2^2)B(p_3^2)],& \\
& O^{\mu \tau}(p_4, p_1, p_2)= \textit{Tr}\gamma^\mu S_F(p_1)  \gamma^\tau S^{-1}_F(p_4) S_F(p_2) & \nonumber \\
&=\frac{1}{[p_1^2A^2(p_1^2)-B^2(p_1^2)][p_2^2A^2(p_2^2)-B^2(p_2^2)]} \times &\nonumber \\
& \bigg[p_1\rho  p_{4\sigma } p_{2 \lambda} M^{\mu \rho \tau \sigma \lambda}A(p_1^2) A(p_2^2)A(p_4^2) + 4 p_{1\rho}p_{2\sigma}(g^{\mu \rho} g^{\tau \sigma} - g^{\mu \tau} g^{ \rho\sigma} +g^{\mu\sigma } g^{ \rho \tau})A(p_1^2) A(p_2^2) B(p_4^2) &\nonumber \\
&- 4 p_{4\rho}p_{2\sigma}(g^{\mu  \tau} g^{\rho \sigma} - g^{\mu \rho} g^{\tau \sigma} + g^{\mu \sigma} g^{\tau  \rho }) B(p_1^2)A(p_2^2)A(p_4^2)   +2i\epsilon^{\mu \tau \rho} p_{2\rho} B(p_1^2)A(p_2^2)B(p_4^2) &\nonumber \\
&-4 p_{1\rho}p_{4\sigma}(g^{\mu \rho} g^{\tau \sigma} - g^{\mu \tau} g^{ \rho\sigma} +g^{\mu\sigma } g^{ \rho \tau}) A(p_1^2)B(p_2^2) A(p_4^2)  +2i\epsilon^{\mu \rho \tau } p_{1\rho}A(p_1^2)  B(p_2^2) B(p_4^2) &\nonumber \\
&-2i\epsilon^{\mu \tau \rho} p_{4\rho}  B(p_1^2)B(p_2^2)A(p_4^2) + 2g^{\mu \tau} B(p_1^2)B(p_2^2)B(p_4^2)\bigg], &  \\
&F^\mu(p_3, p_1,p_2)=\textit{Tr} \gamma^\mu S_F(p_1) S^{-1}_F(p_3)  S_F(p_2) & \nonumber \\
&=\frac{1}{[p_1^2A^2(p_1^2)-B^2(p_1^2)][p_2^2A^2(p_2^2)-B^2(p_2^2)]} \times &\nonumber \\
&  [4 p_{1\rho}p_{3\sigma} p_{2\lambda} (g^{\mu \rho}g^{\sigma \lambda} -g^{\mu \sigma}g^{\rho \lambda} + g^{\mu  \lambda}g^{\rho\sigma})A(p_1^2) A(p_2^2)A(p_3^2)  -2i\epsilon^{\mu \rho \sigma}p_{1\rho} p_{2\sigma}  A(p_1^2) A(p_2^2) B(p_3^2)  +2ip_{3\rho} p_{2\sigma}\epsilon^{\mu \rho \sigma}  B(p_1^2)A(p_2^2)A(p_3^2) &\nonumber \\
&  - 2p_2^\mu B(p_1^2)A(p_2^2)B(p_3^2) + 2ip_{1\rho} p_{3 \sigma}\epsilon^{\mu \rho \sigma}  A(p_1^2)B(p_2^2) A(p_3^2) - 2p_1^\mu  A(p_1^2)  B(p_2^2) B(p_3^2) +2p_3^\mu B(p_1^2)B(p_2^2)A(p_3^2) ],&  \\
 & M^{\mu \nu \rho \tau \lambda}= \textit{Tr}(\gamma^\mu \gamma^\nu \gamma^\rho \gamma^\tau \gamma^\lambda)  =-2ig^{\mu \nu}\epsilon^{\rho \tau \lambda} + 2ig^{\mu \rho}\epsilon^{\nu \tau \lambda}-2ig^{\mu \tau}\epsilon^{\nu \rho \lambda}+i4 \epsilon^{\mu \lambda m}  (g^{\nu \rho}g^{\tau }_m- g^{\nu \tau}g^{\rho  }_m +g^{\nu  }_mg^{  \rho \tau}).& 
\end{flalign} 
\end{widetext}

\nocite{*}

\bibliography{apssamp}

\begin{thebibliography}{99}
\bibitem{ward}J. C. Ward, Phys. Rev. 78 (1950) 182 ;  Y. Takahashi,  II Nuovo Cimento 6(2) (1957) 371-375.
\bibitem{proof}C. Itzykson and Jean-Bernard Zuber, Quantum field theory, McGraw-Hill, New York, 1980.\bibitem{ballchiu}J. S. Ball and Ting-Wai. Chiu, Phys. Rev. D 22 (1980) 2542.
\bibitem{twardys}Y. Takahashi, Quantum Field Theory,  Elsevier Science Publishers, 1986.
\bibitem{tk}Y. Takahashi, Canonical Quantization and Generalized Ward Relations: Foundation of
Nonperturbative Approach, (Print-85-0421 ALBERTA).
\bibitem{tward3}Han-xin He, Phys. Rev. C 63 (2001) 025207.
\bibitem{tward2}Han-xin He, F. C. Khanna and Y. Takahashia, Phys. Lett. B 480 (2000) 222.
\bibitem{tward}Kei-Ichi Kondo, Int. J. Mod. Phys. 12 (1997) 5651. 
\bibitem{penwill}M.R. Pennington and R. Williams, J. Phys. G: Nucl. Part. Phys 32 (2006) 2219.\bibitem{twardzx}Han-xin He, Phys. Rev. D 80 (2009) 016004.
\bibitem{tward8}Han-xin He, Int. J. Mod. Phys. A 22 (2007) 2119.
\bibitem{wt1}A. Bashir and M. R. Pennington, Phys. Rev. D 50 (1994) 7679. P. Maris and C. D. Roberts, Phys. Rev. C 56 (1997) 3369; A. Bashir and A. Raya, Phys. Rev. D 64 (2001) 105001; Ayşe Kızılersü and Michael R. Pennington, Phys. Rev. D 79 (2009) 125020; A. Bashir, R. Bermudez, L. Chang, and C. D. Roberts, Phys. Rev. C 85 (2012) 045205; Ayşe Kızılersü, Tom Sizer, and Anthony G. Williams, Phys. Rev. D 88 (2013) 045008; A. C. Aguilar, J. C. Cardona, M. N. Ferreira and J. Papavassiliou, Phys. Rev. D 96 (2017) 014029.
\bibitem{wt2} Pieter Maris and Peter C. Tandy, Phys. Rev. C 61 (2000) 045202; Williams, R. Eur. Phys. J. A (2015) 51: 57; Mario Mitter, Jan M. Pawlowski and Nils Strodthoff, Phys. Rev. D 91(2015) 054035; Richard Williams, Christian S. Fischer, and Walter Heupel, Phys. Rev. D 93(2016) 034026; Anton K. Cyrol, Mario Mitter, Jan M. Pawlowski, and Nils Strodthoff, Phys. Rev. D 97 (2018) 054006.
\bibitem{qin}Si-Xue Qin, Lei Chang, Yu-Xin Liu, Craig D. Roberts and Sebastian M. Schmidt, Phys.Lett. B722 (2013) 384;
\bibitem{xia}Yong-hui Xia, Hong-tao Feng and Hong-shi Zong, Phys. Rev. D 98 (2018) 074019.
\bibitem{abj1}S.L. Adler, Phys. Rev. 177 (1969) 2426;  J.S. Bell, R. Jackiw, Nuovo Cimento A 60 (1969) 47.
\bibitem{tward4}Wei-Min Sun, Hong-Shi Zong, Xiang-Song Chen and Fan Wang, Phys. Lett. B 569 (2001) 211.
\bibitem{cbl}Cui-Bai Luo, Chen Wu, Song Shi and Hong-Shi Zong, Phys. Lett. B 787 (2018) 39.
\bibitem{wrantner01}W. Rantner and X. G. Wen, Phys. Rev. Lett. 86 (2001) 3871.
\bibitem{PhysRevLett87257003}M. Franz and Z. Te$\check{s}$anovi\'c, Phys. Rev. Lett. 87 (2001) 257003.
\bibitem{PhysRevLett88047006}Igor F. Herbut, Phys. Rev. Lett. 88 (2002) 047006.
\bibitem{spinliquid}Y. Ran, M. Hermele, P. A. Lee, and X. G. Wen, Phys. Rev. Lett. 98 (2007) 117205.
\bibitem{wrantner}W. Rantner and X. G. Wen, Phys. Rev. B 66 (2002) 144501.
\bibitem{khvecomment}D. V. Khveshchenko, Phys. Rev. Lett. 90 (2003) 199701.
\bibitem{khveprb02}D. V. Khveshchenko, Phys. Rev. B 65 (2002) 235111.
\bibitem{khvenpb02}D. V. Khveshchenko, Nucl. Phys. B 642 (2002) 515.
\bibitem{gusunin03}V. P. Gusynin, D. V. Khveshchenko, and M. Reenders, Phys. Rev. B 67 (2003) 115201.\bibitem{franz03}M. Franz, T. Peres-Barnea, D. E. Sheehy, and Z. Tesanovic, Phys. Rev. B 68 (2003) 024508.
\end{thebibliography}

\end{document}